\documentclass[twocolumn,prl,showpacs,preprintnumbers,amsmath,amssymb]{revtex4}

\usepackage{graphicx}
\usepackage{dcolumn}
\usepackage{bm}

\begin{document}

\title{The role of glass dynamics in the anomaly of the dielectric function of solid helium}

\author{Jung-Jung Su$^{1,2}$}
\author{Matthias J. Graf$^1$}
\author{Alexander V. Balatsky$^{1,2}$}
\affiliation{$^1$Theoretical Division, Los Alamos National Laboratory, Los Alamos, New Mexico 87545, USA \\
$^2$Center for Integrated Nanotechnologies, Los Alamos National Laboratory, Los Alamos, New Mexico 87545, USA}
\date{\today}

\begin{abstract}
 We propose that acousto-optical coupling of the electric field to strain fields around defects
in disordered $^4$He is causing an increase of the dielectric function with decreasing temperature due to
the arrested dynamics of defect excitations.
A distribution of such low-energy excitations can be described within the framework of a glass susceptibility of a small volume fraction inside solid $^4$He.
Upon lowering the temperature the relaxation time $\tau(T)$ of defects diverges and
an anomaly occurs in the dielectric function $\epsilon (\omega, T)$
when $\omega \tau(T) \sim 1$. Since $\epsilon (\omega, T)$ satisfies the Kramers-Kronig relation,
we predict an accompanying  peak in the imaginary part of  $\epsilon (\omega, T)$ at the same temperature,
where the largest change in the amplitude has been seen at fixed frequency.
We also discuss recent measurements of the amplitude of the dynamic dielectric function
that indicate  a low-temperature anomaly similar to the one seen in the resonance frequency of the
torsional oscillator and shear modulus experiments.
\end{abstract}

\pacs{67.25.dt, 67.80.B-, 67.80.bd}
\maketitle

The reported anomalies in resonance frequency and dissipation of torsional oscillators
\cite{TO_history}
and in the dynamic shear modulus \cite{SM_Beamish}
at low temperatures  are the subject of intense studies as they were suggested to be the signatures of supersolidity. It is now generally agreed that disorder plays an essential role for observing these effects.
The challenge of unambiguously identifying Bose-Einstein condensation (BEC) in solid $^4$He arises from the fact that crystal defects are required for
supersolidity \cite{THEORIES}.
These defects exhibit their own dynamics and contribute to the observed period of oscillations, shear modulus and specific heat in the same temperature and pressure range, where supersolidity is expected.
Thus any unambiguous identification of a supersolid state   requires a detailed understanding of the behavior of defects.

To study the role of glassy excitations, which make up a small volume fraction of the crystal,
we proposed a phenomenological  framework that captures the dynamics of defects \cite{TO_glass,heatCap_glass}.
It accounts for observed anomalies in torsional oscillator, specific heat, and shear modulus experiments \cite{TO_glass,heatCap_glass,SM_glass}
and can be thought of as a distribution of crystal defects forming two-level-systems (TLS)
\cite{TLS_others}.
Possible candidates for the TLS are groups of atoms near a crystal defect or pinned segments of dislocation lines.
These defects can be removed through annealing or created by applying shear stress, thus drastically changing the mechanical and elastic properties of solid $^4$He \cite{Reppy10}.
A freezing out of these excitations can account for the anomalies by postulating a relaxation time diverging with decreasing temperature. This is typical of glassy systems overcoming an activation barrier
\cite{TO_glass,SM_glass}, which results in visco-elastic behavior,
long relaxation times, hysteresis, as well as effective softening of elastic moduli.

Given the growing experimental evidence for the role of disorder in solid $^4$He,
we ask the question about the effects of a glassy component on the dielectric function at lowest temperatures.
Here we show that  glassy dynamics can affect the strain and position of atoms, which in turn leads to an additional contribution to the  polarization. This contribution causes a decrease of the dielectric function
at high temperatures. Therefore, we propose that the cooperative motion of displaced (out-of-equilibrium) atoms near defects, in the glasslike regions of solid, creates  local strain and thus reduces the polarization through acousto-optical coupling.
Note that this effect is not captured by the standard Clausius-Mossotti equation for dielectrics,
which attributes the change in dielectric function
$\epsilon$ to either a change in polarizability $\alpha$ or in mass density $\rho$:
${(\epsilon-1)}/{(\epsilon+2)} = ({4 \pi}/{3}) ({\alpha \rho}/{M}) $, with molar mass $M$.
In fact, the glass model yields an effect opposite in sign and orders of magnitude bigger than a correction in the
polarizability,  predicted for dipole-induced dipole interactions \cite{Kirkwood}, which was reported a long time ago in $^4$He by Chan et al.\  \cite{Chan77}.

The glass  model describes  the anomaly in $\epsilon(\omega, T)$  due to acousto-optical coupling between strain induced by glassy defects and dielectric polarization. It features:  (i)  a decrease of  $\epsilon(\omega, T)$ at high temperatures;  (ii)  a dissipation peak in the same temperature range where the amplitude of  $\epsilon(\omega, T)$ changes most significantly.  The origin for this dynamic behavior is set by the matching condition $\omega \tau(T) \sim 1$
between the frequency $\omega$ and relaxation time $\tau(T)$.
The key assumption is that a strain field in the vicinity of crystalline defects induces changes in the polarization, since the magnitude of the local electric field sensed by a helium atom is
different from the electric field at equilibrium atomic  position in a crystal free of internal stress.

Very recent  measurements of $\epsilon(\omega, T)$ by Yin and coworkers \cite{Yin11} show the increase of dielectric function at low temperatures. We  illustrate that these data cannot be described by the standard Clausius-Mossotti equation through a change in mass density or polarizability, e.g., due to dipole-induced dipole interactions.
Neither the measured change of the mean mass density $\delta \rho/\rho \sim 10^{-6}$,
nor the predicted correction in polarizability, which actually leads to a decrease of $\epsilon(\omega, T)$ at low temperatures \cite{Chan77, Kirkwood},
can account for the reported change of the dielectric function of order
$\delta \epsilon /\epsilon \sim 10^{-5}$, when
invoking the Clausius-Mossotti equation.

The minimal coupling model between polarization and strain field is obtained by expanding the microscopic polarization, $\bf P^{\rm (micro)}$, around its equilibrium value:
\begin{eqnarray}
{\bf P^{\rm (micro)} (r) \approx P^{\rm (micro)}(R)+ (u \cdot \nabla) \ P^{\rm (micro)}} ,
\end{eqnarray}
where $\bf r=R+u$ is the atom's position, $\bf R$ is its equilibrium position, and $\bf u$ is its displacement.
The  macroscopic polarization $\bf P$ measured in experiment is related to ${\bf P}^{\rm (micro)}$ by the average
over a macroscopic volume element $v$,
${\bf P}=\frac{1}{v}\int_v d {\bf r \ P^{\rm (micro)} (r)}.
$ In the presence of a local strain field one finds in linear order of the displacement
$ {\bf P} \approx \int d {\bf R \ [P^{\rm (micro)}(R)
+ (u \cdot \nabla) \ P^{\rm (micro)}(r)]}.
$ Integration of the first term gives the macroscopic polarization for zero internal strain (a solid in equilibrium), ${\bf P}_0 \equiv 4 \pi \chi_0 {\bf E}$ where $\chi_0$ is the zero-strain susceptibility. The second term can be treated through integration by parts, that is,
$\int d {\bf R \  (u \cdot \nabla) \ P^{\rm (micro)}}=
\int_{S} d A \ {\bf (u \cdot n) \, P^{\rm (micro)}} -
\int d {\bf R \ (\nabla \cdot u)\ P^{\rm (micro)} }$.
The ${\bf u \cdot n}$ term vanishes at the surface of $v$, since the strain field is localized within a macroscopic distance.
The second term,  assuming a smoothly varying strain field, modifies the polarization by
\begin{eqnarray}
\delta {\bf P} \sim e_{ii} \int d {\bf R \ P^{\rm (micro)}(R)}= - e_{ii} \,{\bf P}_0 ,
\end{eqnarray}
with the dilatory strain $e_{ii} \equiv {\bf \nabla \cdot u}$ (we use the notation $e_{ii}\equiv e_{11} + e_{22} + e_{33}$).
This change in local polarization corresponds to a change in dielectric function of
\begin{eqnarray} \label{deltaepsilon}
\delta \epsilon_{ii} \, (\omega, T)
= - 4 \pi \chi_0 \, e_{ii} \, (\omega, T) .
\end{eqnarray}
Note that only the diagonal components of the strain tensor play a role in this leading order expansion.
In principle, the shear strain can couple to the electric field by considering dipole-induced dipole interactions (Van der Waals), which is
a higher order effect. Note that our derivation of Eq.~(\ref{deltaepsilon}) is equivalent up to leading order to the Pockels coefficient for acousto-optical coupling in
isotropic or polycrystalline dielectrics,
$
\delta \epsilon_{ij}= - \epsilon^2 \,[ 2 P_{44} \, e_{ij} + P_{12} \, e_{kk} \delta_{ij}],
$ 
where $P_{kl}$ are the reduced Pockels coefficients (of order unity) \cite{Pockels}.

Next we discuss the origin of the local strain. Depending on the rate of the solidification process of  helium,
various amounts and kinds of defects can be frozen into the crystal leaving behind relics of the history of the phase transition.
We postulate that defects in the glassy regions of solid form the TLS.
The associated strain fields are localized near the TLS and exhibit glassy dynamics, since nearby atoms reshuffle with an intrinsic time delay.
We describe the equation of motion of these atoms within linear response theory by including a back action term \cite{TO_glass,SM_glass}.
In an isotropic or polycrystalline medium the elastic stress tensor
 $\sigma_{ij}^{\rm He} = \lambda_{ijkl} \, \partial u_k/\partial x_{l}$
takes a simplified form with the elastic modulus tensor $\lambda_{ijkl} = \lambda_0 \delta_{ij} \delta_{kl} + \mu_0( \delta_{ik}\delta_{jl} + \delta_{il}\delta_{jk})$.
If the electric field  couples  to local density fluctuations only through dilatory strain,
then the important matrix element is the Lam\'e parameter $\lambda_{0}$
of the purely elastic solid. We write the displacement to an out-of-equilibrium internal force in the presence of the back action as
\begin{eqnarray} \label{eEOM}
\rho \ \partial_{t}^2  \,u_i (t)+ \lambda_{0} \, \partial_i^2 u_i (t)
= f_i^{\rm INT} (t)+ f_i^{\rm BA} (t),
\end{eqnarray}
where $u_i$ is the displacement of an atom in the $i$th direction and
$\rho$ is the mass density.
 $f_i^{\rm INT }$ is the out-of-equilibrium internal force density in the $i$th direction at the defect. $f_i^{\rm BA}$ is the back action force density that describes the time-delayed response of nearby atoms:
\begin{eqnarray} \label{fBA}
f_i^{\rm BA} (t) = \int_{-\infty}^{t} dt' \,{\cal G}(t-t';T) \,\partial_i^2 \, u_i(t') ,
\end{eqnarray}
where ${\cal G}$ is the strength of the back action onto bulk $^4$He.
Integration of Eq.~(\ref{eEOM}) over the $i$th direction and Fourier transformation to frequency domain yields
the dynamic strain due to a local dilatory stress $\sigma_{ii}$:
\begin{eqnarray}
{e}_{ii}(\omega, T) = {\sigma}^{}_{ii}
\left(\lambda_{0}- {\cal G} (\omega, T) \right)^{-1} .
\end{eqnarray}
We assumed that the back action can be described by a distribution $P(t)$ of Debye relaxors,
$
{\cal G} (\omega,T)= \int_0^{\infty}
dt \ P(t)
\,\left[1- {1}/({1- i \omega \,\tau (T) \, t} )\right] ,
 $
with relaxation time $\tau(T)$ of the glassy regions of  solid.
The specific form of $\tau(T)$ can change quantitatively, but not qualitatively, the $T$-dependence of $e_{ii}$.
For simplicity, we choose for ${P(t)}$ the Cole-Cole distribution resulting in
${\cal G}(\omega, T) =  {g_0} \lambda_0/{\big[1-(i \omega \, \tau) ^{\alpha}\big]}$.
Therefore the corresponding dilatory strain of a  distribution of  glassy TLSs becomes
\begin{eqnarray} \label{strain_glass}
{e}_{ii} (\omega, T) &=& e_0 \left(1- {g_0}/\big[{1-(i\omega \tau)^{\alpha}}\big] \right)^{-1} ,
\end{eqnarray}
where $e_0 \equiv \sigma_{ii}/\lambda_0$
at $T=0$
and $g_0$ is the glass parameter, which depends on the TLS density.
From Eq.~(\ref{strain_glass}) and (\ref{deltaepsilon}), we obtain the change in dielectric function due to local strain fluctuations,
\begin{eqnarray}
\delta \epsilon_{ii} &=& -4 \pi \chi_0 \, e_0
\left(1- {g_0}/\big[{1-(i\omega \tau)^{\alpha}}\big] \right)^{-1} .
\end{eqnarray}
At low temperatures, $\tau \rightarrow \infty$ and $e_{ii} \rightarrow e_0$, hence the strain is minimal and the reduction of  dielectric function due to strain fluctuations is small.
At high temperatures, $\tau \rightarrow 0$ and $e_{ii} \rightarrow e_0 (1-g_0)^{-1}$ reaches its maximum resulting in the largest reduction of  strain,  where solid $^4$He is softest.
This is the main result of this study, namely that the dielectric function reflects the arrested dynamics of the glassy components at low temperatures through the acousto-optical coupling.
Within this approximation the maximum size of $\delta \epsilon_{ii}$ is independent of frequency and is controlled by $e_0$ and  $g_0$.

\begin{figure}
\begin{center}
\includegraphics[clip,width=.95\linewidth,angle=0,keepaspectratio]{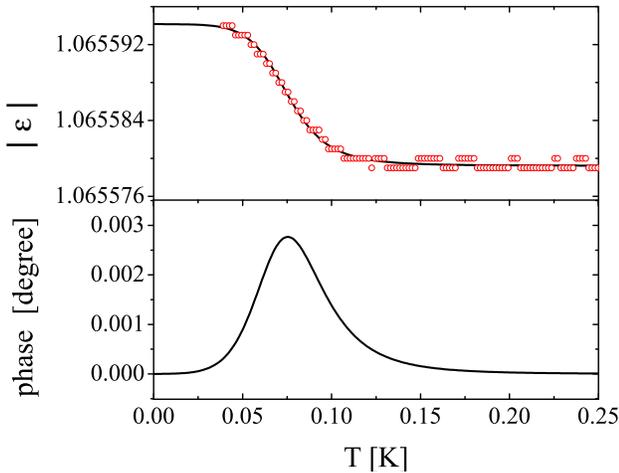}
\end{center}

\vskip-.3cm
\caption{(Color online) Experimental data and calculations of the dielectric function vs.\ temperature.
The red circles are the experimental data of the dielectric function (data by Yin et al.\ \cite{Yin11}).
The black lines are the calculated amplitude and phase lag (dissipation) of $\epsilon(\omega, T)$.
We used parameters $\alpha=1.49$, $e_0=8.88\times 10^{-4}, g=0.21$, $\tau_0=10.4$ ns, $\Delta=1.92$ K, and $T_g=-119$ mK.
}\label{fig:epsilon}
\end{figure}

The results of the acousto-optical model for the dielectric function are shown in Fig.~\ref{fig:epsilon}.
We obtain excellent agreement with the measurements by Yin  et al.\ \cite{Yin11} for an applied alternating voltage at 500 Hz.
Additionally, we predict an accompanying dissipation peak in the dielectric function or a phase lag
$\phi=arg( \epsilon )$ between the real and imaginary part of $\epsilon(\omega, T)$. Experimental confirmation of the dissipation peak will constitute an important test of the glassy nature and the acousto-optical coupling of defects in solid helium.
In our calculation, we assumed the Vogel-Fulcher-Tammann (VFT) form for the glassy relaxation time: $\tau(T) = \tau_0 \,e^{\Delta/(T-T_g)}$ for $T>T_g$,  and $\tau(T)=\infty$ for $T \le T_g$,
which describes non-Arrhenius thermal activation processes for $T_g\neq 0$. Here $\tau_0$ is the attempt time and $\Delta$ is the mean activation energy of a TLS. In our fit, we are not biasing  the ideal glass transition temperature $T_g$ to be positive. We find a negative $T_g$, which means no real phase transition occurs at finite temperatures, but rather a crossover. This can be due to inherent quantum fluctuations originating from the large zero-point motion of helium atoms, which suppress the onset of a lower phase transition.

\begin{figure}
\begin{center}
\includegraphics[width=0.90\linewidth,angle=0,keepaspectratio]{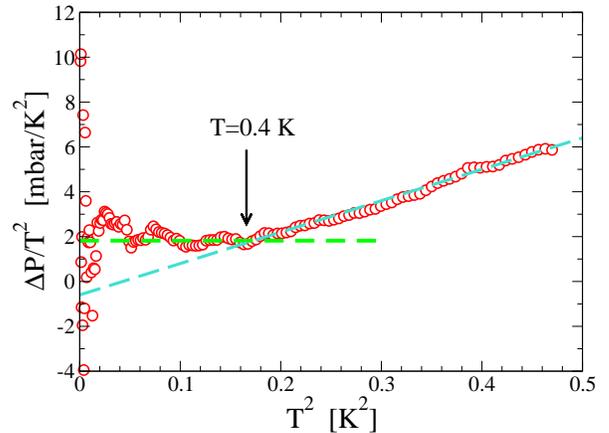}
\end{center}
\vskip-.3cm

\caption{(Color online) Low-temperature pressure deviation from lattice contribution of Debye solid (data by Yin et al.\ \cite{Yin11}). The intercept of $\Delta P/T^2$ vs.\ $T^2$ extrapolated from low-temperature data points is in agreement with a glassy contribution from TLS on the order of 100 ppm.
The arrow indicates the onset of  deviation from the Debye solid. The large scatter of data points at lowest temperatures is due to the subtraction of a constant pressure $P_0$. Dashed lines are guides to the eyes.
}\label{fig:PT}
\end{figure}
A discussion of the glass model in the context of the dielectric function requires a justification for the magnitude of the
elasto-optical coupling to strain.
The existence of local strain in solid $^4$He
is inferred from  measured pressure gradients.  We obtain a rough estimate, for example, from the mass flow measurements in bulk solid $^4$He \cite{Hallock}, where a pressure difference of $\Delta P_{L} \sim 0.1$ bar across a centimeter-sized pressure cell is reported. The estimated local strain, with a bulk modulus $B=320$ bar, is accordingly  $\Delta P_{L}/B = 3 \times 10^{-4}$. This is consistent with the value we used for the fit in Fig.~1,  namely $e_0 = 8.88 \times 10^{-4}$.
In fact the local strain may have been even larger in the experiments by Yin due to the more complicated geometry of the measurement apparatus.

Since the  acousto-optical coupling model depends on the presence of defects, an important question to ask is what is the quality of the helium crystals. For that reason we are re-analyzing pressure measurements  by Yin \cite{Yin11}.
The glassy contribution of the TLS  to the bulk pressure through $P(T) =  P_0 + a_{gl} T^2 + a_{ph} T^4$,
where  $P_0$ is the crystal pressure of bulk solid at $T=0$, while coefficients  $a_{gl}$ and $a_{ph}$ correspond to the contributions of the TLS glass and phonons, respectively.
In Fig.~\ref{fig:PT} we  show the temperature dependence of the deviation $\Delta P \equiv P(T)-P_0$.
A large glassy contribution can be seen at low temperatures similar to reports by others
\cite{Rudavskii}.
The $P(T)$ curve in Yin's measurement clearly deviates from a purely Debye lattice behavior at around $T = 0.4$ K with a large positive intercept corresponding to the order of 100 ppm of TLS \cite{heatCap_glass}. This number is roughly five times larger than the most disordered sample in Lin's \cite{Lin09} specific heat experiments on ultrapure $^4$He with less than 1 ppb of $^3$He impurities grown over four hours using the blocked capillary method \cite{Lin09,heatCap_glass}.
Clearly crystals grown by Yin and coworkers are disordered with sufficiently many TLS defects to support
centers of local strain fields.

So far we assumed that both local and global stress are constant at low temperatures.
From Fig.~\ref{fig:PT} we can read off that the global pressure change between 300 mK and 40 mK is less than $\Delta P_T=0.18$ mbar. This is more than three orders of magnitude smaller than the local stress $\sigma_L = 8.88 \times 10^{-4} \times 320 \ {\rm bar} \sim 250$ mbar inferred from the dilatory strain $e_0$ used in the fit, as well as the static pressure difference  $\Delta P_L = 0.1$ bar measured at two pressure gauges in the mass flow experiment \cite{Hallock}.
Taking all these observations together, we find that the change of the dielectric function based on global
density changes in the Clausius-Mossotti equation is negligible. The corresponding density change is
$\Delta\rho/\rho = \Delta P_T/B < 10^{-6}$, leading to a change in the dielectric function of
$\delta \epsilon \approx (\epsilon-1) \Delta \rho /\rho = 0.065 \Delta P_T/B < 10^{-7}$, which is more than two orders of magnitude too small to account for the observed effect $\sim10^{-5}$.

In conclusion, we have shown that the arrested glass dynamics causes low-temperature anomalies in strained solid $^4$He through the acousto-optical coupling. 
We argued that the temperature behavior of the dielectric function is coupled  to local strain fields near crystal defects.
Thus it records the glassy dynamics and  concomitant freeze-out of TLS excitations, which also leads to a stiffening of solid with decreasing temperature. This effect is not captured by the standard Clausius-Mossotti relation, which attributes dielectric function changes to a change in mass density or polarizability of the helium atom.
An important consequence of the phenomenological glass susceptibility is the decrease of dielectric function at high temperatures, accompanied by a broad dissipation peak that  can be measured  by imaginary part of dielectric function. We hypothesize that the cooperative motion of atoms forming the TLS along dislocation segments is the relevant process contributing to the reported anomaly.
In our model, both the change in $\epsilon(\omega, T)$ and the dissipation peak are independent of  applied frequency.
Since  the glass parameter $g_0$ of the back action depends on the concentration of the TLS, we predict that the change in dielectric function will be larger in quench cooled or shear stressed samples.

We acknowledge fruitful discussions with Z. Nussinov and J. C. Davis. We are grateful
to L. Yin and N. Sullivan for explaining their experiments and sharing their data.
This work was supported by the
U.S.\ DOE at Los Alamos National Laboratory under Contract No.~DE-AC52-06NA25396 through the LDRD program
and the Center for Integrated Nanotechnologies, an Office of Basic Energy Sciences user facility.

\end{document}